\documentclass[twocolumn]{aastex61}
\pdfoutput=1 %for arXiv submission
\usepackage{amsmath,amstext}
\usepackage[T1]{fontenc}
\usepackage{apjfonts} 
\usepackage[figure,figure*]{hypcap}
\usepackage[hang,flushmargin]{footmisc} 
\usepackage{lineno}

\usepackage{amsmath}
\usepackage[caption=false]{subfig}
\usepackage{graphicx}

\usepackage{hyperref}
\usepackage{url}

\usepackage{enumitem}
\setlist[enumerate]{label*=\arabic*.}

\newcommand{\apx}{\ensuremath{\sim}}

\shorttitle{GW170817 EM Counterpart Paper II}
\shortauthors{Cowperthwaite et al.}
\begin{document}

\title{The Electromagnetic Counterpart of the Binary Neutron Star Merger LIGO/VIRGO GW170817. II. UV, Optical, and Near-IR Light Curves and Comparison to Kilonova Models}

%% Primary Authors
\author{P.~S.~Cowperthwaite}
\affiliation{Harvard-Smithsonian Center for Astrophysics, 60 Garden Street, Cambridge, Massachusetts 02138, USA}
\author{E.~Berger}
\affiliation{Harvard-Smithsonian Center for Astrophysics, 60 Garden Street, Cambridge, Massachusetts 02138, USA}
\author{V.~A.~Villar}
\affiliation{Harvard-Smithsonian Center for Astrophysics, 60 Garden Street, Cambridge, Massachusetts 02138, USA}
\author{B.~D.~Metzger}
\affiliation{Department of Physics and Columbia Astrophysics Laboratory, Columbia University, New York, NY 10027, USA}
\author{M.~Nicholl}
\affiliation{Harvard-Smithsonian Center for Astrophysics, 60 Garden Street, Cambridge, Massachusetts 02138, USA}
\author{R.~Chornock}
\affiliation{Astrophysical Institute, Department of Physics and Astronomy, 251B Clippinger Lab, Ohio University, Athens, OH 45701, USA}
\author{P.~K.~Blanchard}
\affiliation{Harvard-Smithsonian Center for Astrophysics, 60 Garden Street, Cambridge, Massachusetts 02138, USA}
\author{W.~Fong}
\affiliation{CIERA and Department of Physics and Astronomy, Northwestern University, Evanston, IL 60208}
\author{R.~Margutti}
\affiliation{CIERA and Department of Physics and Astronomy, Northwestern University, Evanston, IL 60208}
\author{M.~Soares-Santos}
\affiliation{Department of Physics, Brandeis University, Waltham, MA 02454, USA}
\affiliation{Fermi National Accelerator Laboratory, P. O. Box 500, Batavia, IL 60510, USA}

%% Opt-In (Alphabetical)
% ...

\begin{abstract}
We present UV, optical, and NIR photometry of the first electromagnetic counterpart to a gravitational wave source from Advanced LIGO/Virgo, the binary neutron star merger GW170817. Our data set extends from the discovery of the optical counterpart at $0.47$ days to $18.5$ days post-merger, and includes observations with the Dark Energy Camera (DECam), Gemini-South/FLAMINGOS-2 (GS/F2), and the {\it Hubble Space Telescope} ({\it HST}). The spectral energy distribution (SED) inferred from this photometry at $0.6$ days is well described by a blackbody model with $T\approx 8300$ K, a radius of $R\approx 4.5\times 10^{14}$ cm (corresponding to an expansion velocity of $v\approx 0.3c$), and a bolometric luminosity of $L_{\rm bol}\approx 5\times10^{41}$ erg s$^{-1}$. At $1.5$ days we find a multi-component SED across the optical and NIR, and subsequently we observe rapid fading in the UV and blue optical bands and significant reddening of the optical/NIR colors. Modeling the entire data set we find that models with heating from radioactive decay of $^{56}$Ni, or those with only a single component of opacity from $r$-process elements, fail to capture the rapid optical decline and red optical/NIR colors. Instead, models with two components consistent with lanthanide-poor and lanthanide-rich ejecta provide a good fit to the data; the resulting ``blue'' component has $M_\mathrm{ej}^\mathrm{blue}\approx 0.01$ M$_\odot$ and $v_\mathrm{ej}^\mathrm{blue}\approx 0.3$c, and the ``red'' component has $M_\mathrm{ej}^\mathrm{red}\approx 0.04$ M$_\odot$ and $v_\mathrm{ej}^\mathrm{red}\approx 0.1$c. These ejecta masses are broadly consistent with the estimated $r$-process production rate required to explain the Milky Way $r$-process abundances, providing the first evidence that BNS mergers can be a dominant site of $r$-process enrichment.

\end{abstract}


\section{Introduction}
\label{sec:intro}
The era of gravitational wave (GW) astronomy began on 2015 September 14 when the Advanced Laser Interferometer Gravitational Wave Observatory (LIGO) made the first direct detection of gravitational waves, resulting from the merger of a stellar mass binary black hole (BBH; GW150914; \citealt{ligo1}). LIGO has since announced the detection of three additional BBH events \citep{ligo2,ligo3,ligo4}.  There are currently no robust theoretical predictions for electromagnetic (EM) emission associated with such mergers. 

By contrast, mergers involving at least one neutron star can produce a wide range of EM signals, spanning from gamma-rays to radio (e.g., \citealt{Metzger2012}). In the optical/NIR bands, the most promising counterpart is the kilonova (KN), a roughly isotropic thermal transient powered by the radioactive decay of rapid neutron capture ($r$-process) elements synthesized in the merger ejecta (\citealt{li1998,Metzger2010,Roberts2011,Metzger2012,Barnes2013,Tanaka2013}).    
The properties of the KN emission (luminosity, timescale, spectral peak) depend sensitively on the ejecta composition.  For ejecta containing Fe-group or light $r$-process nuclei with atomic mass number $A\lesssim 140$, the KN emission is expected to peak at optical wavelengths at a luminosity $L_{\rm p}\sim 10^{41}-10^{42}$ erg s$^{-1}$ on a short timescale of $t_{\rm p}\sim 1$ day (a so-called ``blue'' KN; \citealt{Metzger2010,Roberts2011,Metzger2014}).  By contrast, for ejecta containing heavier lanthanide elements ($A\gtrsim 140$) the emission is predicted to peak at NIR wavelengths with $L_{\rm p} \sim 10^{40}-10^{41}$ erg s$^{-1}$ over a longer timescale of $t_{\rm p}\sim 1$ week (a so-called ``red'' KN; \citealt{Barnes2013,Kasen2013,Tanaka2013}).

The first direct detection of gravitational waves from the inspiral and merger of a binary neutron star (BNS) was made on 2017 August 17 (GW170817; \citealt{ALVdetection,ALVSkymapgcn,ALVgcn}). This source was coincident with a short burst of Gamma-rays detected by both {\it Fermi}/GBM and {\it INTEGRAL} (GRB\,170817A; \citealt{GBMgcn1,GBMdetection,GBMgcn3,INTEGRALdetection,INTEGRALgcn,GBMgcn2}). Rapid optical follow-up by our DECam program \citep{DECamref}, starting just 11.4 hours after the GW trigger, led to the discovery of an associated optical counterpart in the nearby ($d\approx 39.5$ Mpc; \citealt{NGC4993dist}) galaxy NGC\,4993 \citep{DECAMgcn,DECampaper1}. This optical source was independently discovered by several groups \citep{GWEMcapstone}, and first announced as SSS17a by \cite{SWOPEpaper,SWOPEgcn}. The source has also been independently named DLT17ck \citep{DLT40paper,DLT40gcn}, and AT2017gfo.

Here we present rapid-cadence UV, optical, and NIR observations spanning from the time of discovery to 18.5 days post-merger.  We construct well-sampled light curves and SEDs using data from DECam along with {\it Swift}, GS/F2, and {\it HST}. We show that the data cannot be fit by a model with heating from $^{56}$Ni radioactive decay and Fe-peak opacities (as in normal supernovae), but instead requires heating from $r$-process nuclei and at least two components consistent with lanthanide-poor and lanthanide-rich opacities. We further use the data to determine the ejecta masses and velocities for each component.

All magnitudes presented in this work are given in the AB system and corrected for Galactic reddening with $E(B-V)=0.105$\footnote{This is computed from \url{http://irsa.ipac.caltech.edu/applications/DUST/} using the coordinate transients in \cite{DECampaper1}.} and applying the calibration of \cite{schlafly11}. We assume a negligible reddening contribution from the host \citep{DECamPaper8}. 

%Cosmological calculations are performed using the cosmological parameters $H_0 = 67.7$ km s$^{-1}$ Mpc$^{-1}$, $\Omega_M = 0.307$, and $\Omega_{\Lambda} = 0.691$ \citep{planck15}.

\begin{figure*}[!t]
\begin{center}
\hspace*{-0.1in} 
\scalebox{1.}
{\includegraphics[width=1.\textwidth]{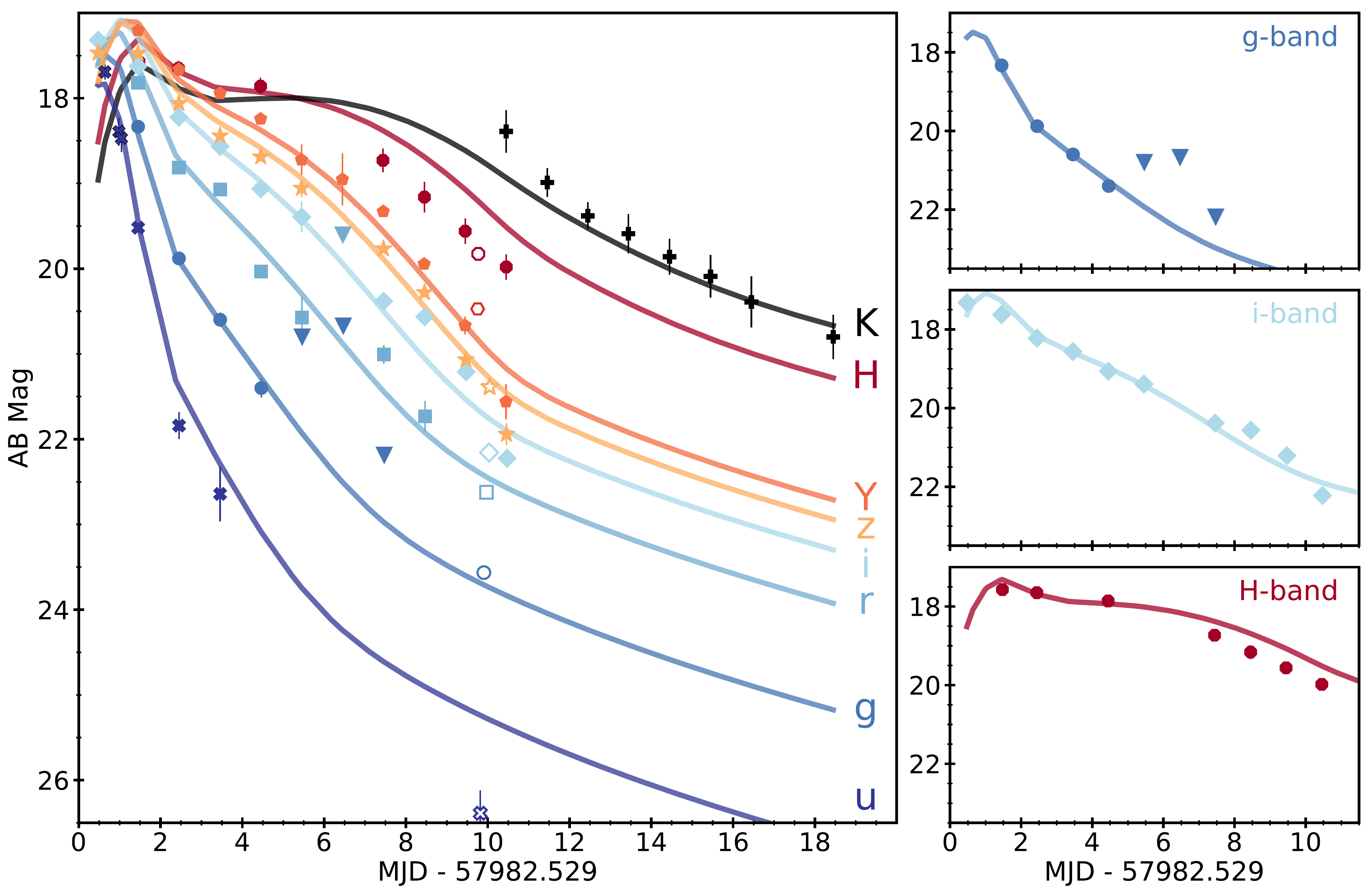}}
\caption{UV, optical, and NIR Light curves of the counterpart of GW170817. The two-component model for $r$-process heating and opacities (\autoref{sec:models}) is shown as solid lines. The right panels focus on the $g$ (top), $i$ (middle), and $H$-band photometry (bottom), over the first 10 nights. Triangles represent 3$\sigma$ upper limits. Error bars are given at the $1\sigma$ level in all panels, but may be smaller than the points.}
\label{fig:lc_good}
\end{center}
\end{figure*}

\section{Observations and Data Analysis}
\label{sec:observations}

A summary of the observations and photometry described in this section is available in \autoref{tab:phot}.

\subsection{DECam}
\label{sec:data}
We processed all DECam images using the {\tt photpipe} pipeline (e.g., \citealt{rest05,rest14}), to perform single-epoch image processing and image subtraction using the {\tt hotpants} software package \citep{becker15}. Point spread function (PSF) photometry was performed on the subtracted images using an implementation of {\tt DoPhot} optimized for difference images \citep{schechter93}. We performed astrometric and photometric calibration relative to the Pan-STARRS1/$3\pi$ catalog (PS1/$3\pi$; \citealt{PS1ref}), with appropriate corrections between magnitude systems \citep{scolnic15}. The typical calibration error is on the order of $\approx3\%$. Image subtraction was performed using stacked images from the PS1/$3\pi$ survey as reference images for $gr$-band. DECam images from 2017 August 25 and 2017 August 31 were used as reference images for $u$-band and $izY$-band, respectively, after the transient had faded away.

\subsection{HST}
\label{sec:data_HST} 
We obtained \textit{HST} Target-of-Opportunity observations of GW170817 on 2017 August 27.28 (9.8 days post-trigger) UT using ACS/WFC with the F475W, F625W, F775W, and F850LP filters, WFC3/UVIS with the F336W filter, and WFC3/IR with the F160W and F110W filters (PID: 15329; PI: Berger).  We retrieved the calibrated data from the Mikulski Archive for Space Telescopes (MAST) and used the {\tt DrizzlePac}\footnote{\url{http://drizzlepac.stsci.edu/}} software package to create final drizzled images from the individual dithered observations in each filter.  We used the {\tt astrodrizzle} task to correct for optical distortion and improve the resolution from that sampled by the instrumental PSF. We measure the flux of the optical counterpart by fitting a model PSF, constructed from multiple stars in each image, using a custom Python wrapper for {\tt DAOPhot} \citep{Stetson87}. We remove contaminating flux from the host galaxy at the transient location using local background subtraction. After subtraction, the typical contribution from the host flux is $\lesssim5\%$. We calibrate the photometry for each image using the zeropoints provided by the HST analysis team\footnote{\url{http://www.stsci.edu/hst/acs/analysis/zeropoints}}.

\subsection{GS/F2}
\label{sec:data_GSF2}
We obtained several epochs of $HK_s$ band photometry using FLAMINGOS-2 on the Gemini-South 8~m telescope \citep{GSF2ref2} starting on 2017 August 19.00 (1.47 days post-merger). We processed the images using standard procedures in the {\tt gemini} IRAF\footnote{IRAF is distributed by the National Optical Astronomy Observatory, which is operated by the Association of Universities for Research in Astronomy (AURA) under a cooperative agreement with the National Science Foundation.} package.  We created an average sky exposure from the individual dithered frames and then scaled and subtracted from each science image prior to registration and combination of the images. We perform PSF photometry using field stars and host galaxy subtraction as described in \autoref{sec:data_HST}, and calibrate the photometry relative to the 2MASS point source catalog \footnote{\url{https://www.ipac.caltech.edu/2mass/}}.

\subsection{Swift/UVOT}
\label{sec:data_UVOT}
The UVOT instrument on-board {\it Swift} \citep{Gehrels04, Roming05} began observing the field of the optical counterpart on 2017 August 18.167 UT with the {\it U}, {\it W1}, {\it W2}, and {\it M2} filters \citep{SWIFTgcn2,SWIFTgcn1,SWIFTpaper}.  We use the latest {\tt HEAsoft} release (v6.22) with the corresponding calibration files and updated zero-points to independently analyze the data. We perform photometry in a $3\arcsec$ photometric aperture to minimize the contamination from host galaxy light, following the prescriptions by \citet{Brown09}. We estimate and subtract the contribution from host galaxy light using deep UVOT observations acquired at later times, when the UV emission from the transient was no longer present in the images ({\it Swift} ID 07012979003). The systematic effect from the host light contamination is $\approx 3\%$ (see e.g., \citealt{Brown09}).

\section{Light Curves and Spectral Energy Distributions}
\label{sec:analysis}

\begin{figure}[!t]
\begin{center}
\hspace*{-0.1in} 
\scalebox{1.}
{\includegraphics[width=0.5\textwidth]{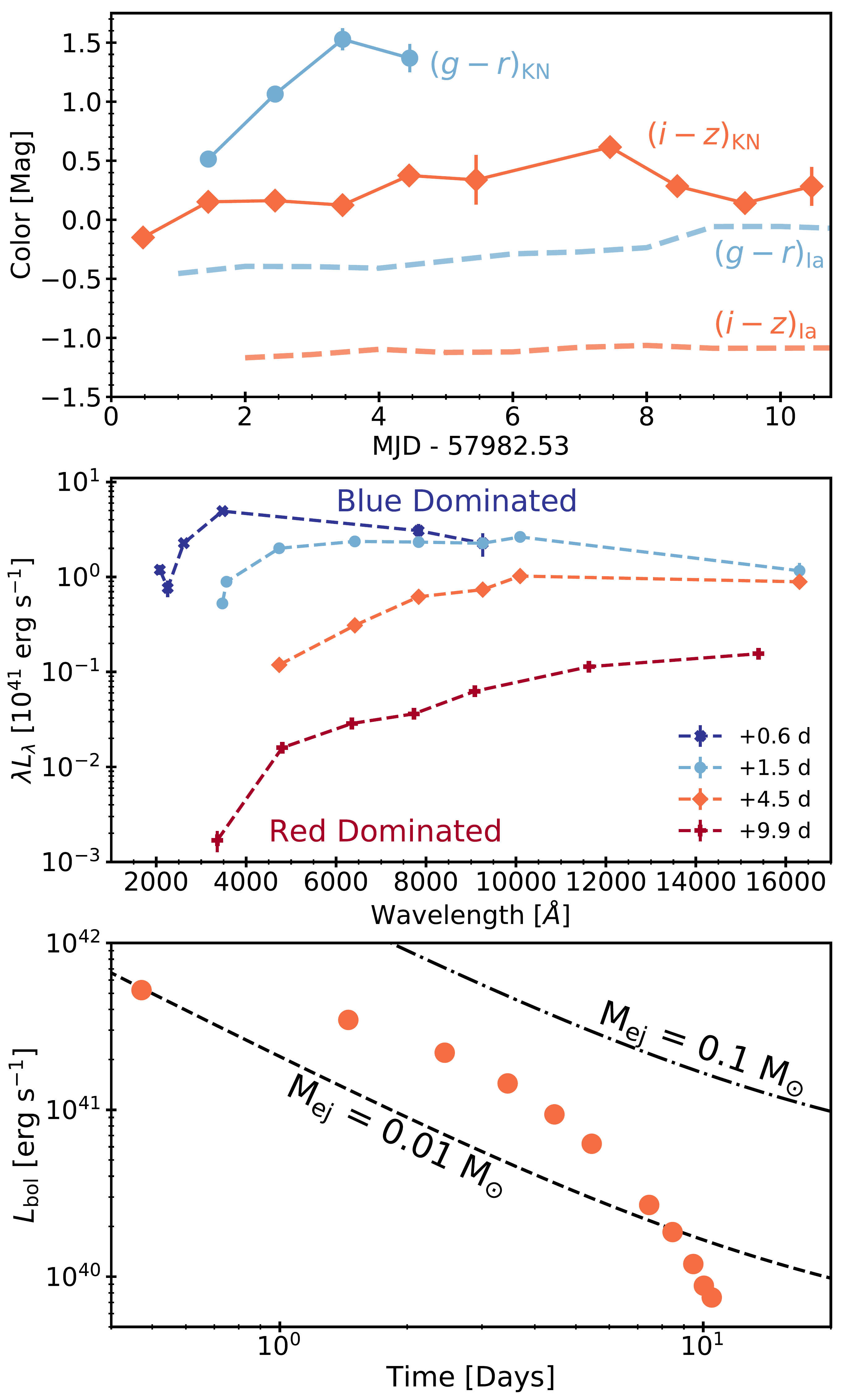}}
\caption{{\it Top:} Optical colors from DECam observations as a function of time.  We observed rapid and early reddening in $g-r$ compared to the relatively flat but red $i-z$ colors. Also shown are template Ia SN colors relative to explosion for comparison \citep{nugent2002k}. {\it Middle:} SEDs at four representative epochs (assuming isotropic emission). The transition from a blue dominated spectrum at early times to a spectrum dominated by a red component at late times is clearly visible. {\it Bottom:} Bolometric light curve spanning $ugrizYH$. Expected values for $r$-process heating from \cite{Metzger2010} are shown for comparison, indicating that the observed emission requires ${\rm few}\times 10^{-2}$ M$_\odot$ of $r$-process ejecta. Error bars are given at the $1\sigma$ level in all panels, but may be smaller than the points.}
\label{fig:sed}
\end{center}
\end{figure}

\subsection{Light Curves}
\label{sec:lc}

Our UV/optical/NIR light curves are shown in \autoref{fig:lc_good}. The data span from 0.47 to 18.5 days post merger, with bluer bands fading below the detection limits at earlier times.  The light curve coverage was truncated by the proximity of the source to the Sun.  We first note that the light curves are not well described by a power law, indicating minimal contribution from a GRB optical afterglow over the timescale of our observations. This is consistent with modeling of the afterglow based on X-ray and radio observations \citep{DECamPaper6,DECamPaper5}.

The light curves exhibit a rapid decline in the bluest bands ($ug$), an intermediate decline rate in the red optical bands ($rizY$), and a shallow decline in the NIR ($HK_s$).  However, while the $u$- and $g$-band light curves decline by $\approx 2$ mag d$^{-1}$ starting with the earliest observations, the redder optical bands exhibit a more complex behavior: they exhibit a comparatively slow decline ($\approx0.3$ mag d$^{-1}$) over the first 1.5 days, develop a shoulder at about 4 days, and subsequently begin to decline at about 8 days.

We find a similar rapid evolution in the colors of the transient (\autoref{fig:sed}).  In particular, the $u-g$ and $g-r$ colors become redder by about 1 mag between about 1.5 and 3.5 days.  The colors in the redder optical bands exhibit slower evolution, with $r-i\approx 0.5-1$ mag, $i-z\approx 0-0.5$ mag, and $z-Y\approx 0.3$ mag.  These colors are significantly redder than those of known supernovae near explosion (e.g., \citealt{folatelli2009,bianco2014,galbany2016}).

\subsection{Spectral Energy Distribution}
\label{sec:SED}

We construct SEDs from photometry at several epochs from about 0.6 to 10 days post-merger (\autoref{fig:sed}). The SEDs exhibit rapid evolution from an initial peak at \apx3500 $\mbox{\AA}$ to a final peak at $\gtrsim 15,000$~$\mbox{\AA}$ by 10 days.  Moreover, the SED at 1.5 days appears to consist of two components, as indicated by the changing slope in the NIR emission. The same rapid evolution and structure are apparent in the optical and NIR spectra at comparable epochs \citep{DECamPaper4,DECamPaper3}.

The SED at 0.6 days is well described by a blackbody with $T\apx 8300$ K and $R\apx 4.5\times 10^{14}$ cm, corresponding to an expansion velocity of $v\apx 0.3c$. This is somewhat larger than the velocities observed in broad-lined Type Ic SNe (for which $v\approx 0.1c$; \citealt{modjaz2016}), but is consistent with expectations for ejecta resulting from a BNS merger \citep{Metzger2017}. The SEDs at later times are not well described by a blackbody curve, instead exhibiting strong flux suppression at blue wavelengths that leads to a spectrum with a sharper peak than a blackbody.  This behavior is also present in our optical spectra \citep{DECamPaper3}.

\subsection{Bolometric Light Curve}
\label{sec:lc_bol}

We construct a bolometric light curve from the $ugrizYH$ data spanning to 11 days. We fit the time evolution in each band independently with a linear model and interpolate the magnitudes to a common grid of times. The bolometric luminosity is determined using the integrated total flux at each time step; see \autoref{fig:sed}. The peak bolometric luminosity of $\apx5\times10^{41}$ ergs s$^{-1}$ at 0.6 days is broadly consistent with the luminosity predicted for $r$-process heating by a ${\rm few}\times 10^{-2}$ M$_{\odot}$ ejecta, similar to the original predictions of \cite{Metzger2010} for blue KN emission from Fe-opacity ejecta. The total radiated energy during the first \apx10 days is $\approx10^{47}$~ergs.

\subsection{Qualitative Comparisons to Kilonova Emission}
\label{sec:kn_compare}
There are several lines of preliminary evidence that suggest the optical counterpart is an $r$-process powered kilonova before exploring detailed models. The presence of an initially blue SED, transitioning to a multi-component SED, and finally to a red SED, is strongly suggestive of both blue and red kilonova emission, consistent with  lanthanide-poor and rich ejecta components, respectively \citep{Metzger2010,Barnes2013,Tanaka2013,Metzger2014,kasen15,Wollaeger2017}. Furthermore, the deviations from a pure blackbody spectrum at late times are indicative of the strong UV line blanketing expected for lanthanide-rich material, lending further evidence to the existence of a red kilonova component. This behavior is also seen in optical/NIR spectra of the transient \citep{DECamPaper4,DECamPaper3}.

The fact that this red component does not initially obscure the emission from a blue component suggests that we require two separate emitting regions with distinct sources of ejecta. If the KN outflow is quasi-spherical, then the blue component must reside {\it outside} the material with red emission.  Alternatively, if the outflow is not spherically symmetric, the blue and red ejecta should occupy distinct portions of the outflowing solid angle. This feature is suggested in several models that consider lanthanide-rich material ejected in the equatorial plane while the lanthanide-poor material is ejected from the polar regions \citep{kasen15,Metzger2017}. 

\startlongtable
\begin{deluxetable*}{cccccccccccc}
\tabletypesize{\footnotesize}
\tablecolumns{12}
\tablewidth{0pt}
\tablecaption{Kilonova Model Fits
	          \label{tab:KN}}
\tablehead{
\colhead{Model} & 
\colhead{$M_{\rm ej}^{\rm blue}$} & 
\colhead{$v_{\rm ej}^{\rm blue}$} &
\colhead{$\kappa^{\rm blue}$} &
\colhead{$M_{\rm ej}^{\rm purple}$} & 
\colhead{$v_{\rm ej}^{\rm purple}$} &
\colhead{$\kappa^{\rm purple}$} &
\colhead{$M_{\rm ej}^{\rm red}$} & 
\colhead{$v_{\rm ej}^{\rm red}$} &
\colhead{$\kappa^{\rm red}$} &
\colhead{$f^{\rm Ni}$} &
\colhead{WAIC} \\
\colhead{} & 
\colhead{($\text{M}_{\odot}$)} & 
\colhead{(c)} & 
\colhead{($\text{cm}^2 \text{g}^{-1}$)} & 
\colhead{($\text{M}_{\odot}$)} & 
\colhead{(c)} & 
\colhead{($\text{cm}^2 \text{g}^{-1}$)} & 
\colhead{($\text{M}_{\odot}$)} & 
\colhead{(c)} & 
\colhead{($\text{cm}^2 \text{g}^{-1}$)} & 
\colhead{}&
\colhead{}
}
\startdata
2-Comp & $0.014^{+0.002}_{-0.001}$ & $0.266^{+0.007}_{-0.002}$ & (0.5) & - & - & - & $0.036^{+0.001}_{-0.002}$ & $0.123^{+0.012}_{-0.014}$ & $3.349^{+0.364}_{-0.337}$  & - & -102 \\
3-Comp & $0.014^{+0.002}_{-0.001}$ & $0.267^{+0.006}_{-0.011}$ & (0.5) & $0.034^{+0.002}_{-0.002}$ & $0.110^{+0.011}_{-0.010}$ & (3.0) & $0.010^{+0.002}_{-0.001}$ & $0.160^{+0.030}_{-0.025}$ & (10.0) & - &  -106 \\
\hline
$^{56}$Ni & $0.008^{+0.007}_{-0.001}$ & $0.260^{+0.034}_{-0.031}$ & (0.1) & - & - & - & - & - & -  & $0.749^{+0.214}_{-0.203}$ & 17\\
Blue & $0.032^{+0.002}_{-0.004}$ & $0.180^{+0.002}_{-0.002}$ & (0.1) & - & - & - & - & - & -  & - & 17\\
Red & - & - & - & - & - & - & $0.026^{+0.010}_{-0.008}$ & $0.271^{+0.008}_{-0.002}$ & (10) & - & 153\\
1-Comp & - & - & - & $0.040^{+0.002}_{-0.007}$ & $0.274^{+0.007}_{-0.093}$ & $0.817^{+0.146}_{-0.135}$  & - & - & - & - &  11
\enddata
\tablecomments{Model parameters and WAIC scores. Numbers in parentheses indicate fixed parameters of the model. The errors represent the 1$\sigma$ confidence interval. Both the 2-component (``2-Comp'') and 3-component (``3-Comp'') models have significantly smaller WAIC scores (indicating better fits) compared to the four single-component models. }
\end{deluxetable*}

\section{Kilonova Modeling}
\label{sec:models}

We test the conjecture that the UV/optical/NIR transient is an $r$-process kilonova by fitting several isotropic, one-zone, gray opacity models to the light curves. For each model, we assume a blackbody SED which evolves assuming a constant ejecta velocity until it has reached a minimum temperature, at which point the photosphere has receded into the ejecta and the temperature no longer evolves. A similar temperature ``floor'' is predicted in \cite{Barnes2013}, and we include this minimum temperature as a fitted parameter.  We additionally fit for a ``scatter'' term, added in quadrature to all photometric errors, which roughly accounts for additional systematic uncertainties that are not included in our model. 

We use {\tt MOSFiT}\footnote{\url{https://github.com/guillochon/MOSFiT}} \citep{mosfitpaper,nicholl2017magnetar}, an open source light curve fitting tool that utilizes a Markov Chain Monte Carlo (MCMC) to sample the model posterior. For each model, we ensure convergence by enforcing a Gelman-Rubin statistic $<1.1$ \citep{gelman1992inference}.  We compare models using the Watanabe-Akaike Information Criteria (WAIC, \citealt{watanabe2010asymptotic,gelman2014understanding}), which accounts for both the likelihood score and number of fitted parameters. The best fit parameters, uncertainties, and WAIC scores for each model are provided in Table~\ref{tab:KN}.

We first attempt a simple supernova model, namely heating by the radioactive decay of $^{56}$Ni and Fe-peak opacity of $\kappa = 0.1$ cm$^{2}$ g$^{-1}$ (see \citealt{villar2017}).  The model parameters are the ejecta mass and velocity, and the $^{56}$Ni mass fraction in the ejecta (as well as the temperature floor and scatter).  The best-fit model has $M_{\rm ej}\approx 0.01$ M$_\odot$, $v_{\rm ej}\approx 0.26$c, and $f^{\rm Ni}\approx 0.75$.  The parameters are comparable to those we inferred from black body fits to the flux and SEDs in the previous section, but the overall fit is poor.  In particular, this model severely underestimates the NIR light curves, while the high $^{56}$Ni fraction is inconsistent with the optical spectra \citep{DECamPaper3}.  We therefore conclude that the transient is not powered by the radioactive decay of $^{56}$Ni.

\begin{figure*}
\begin{center}
\hspace*{-0.1in} 
\scalebox{1.}
{\includegraphics[width=1.\textwidth]{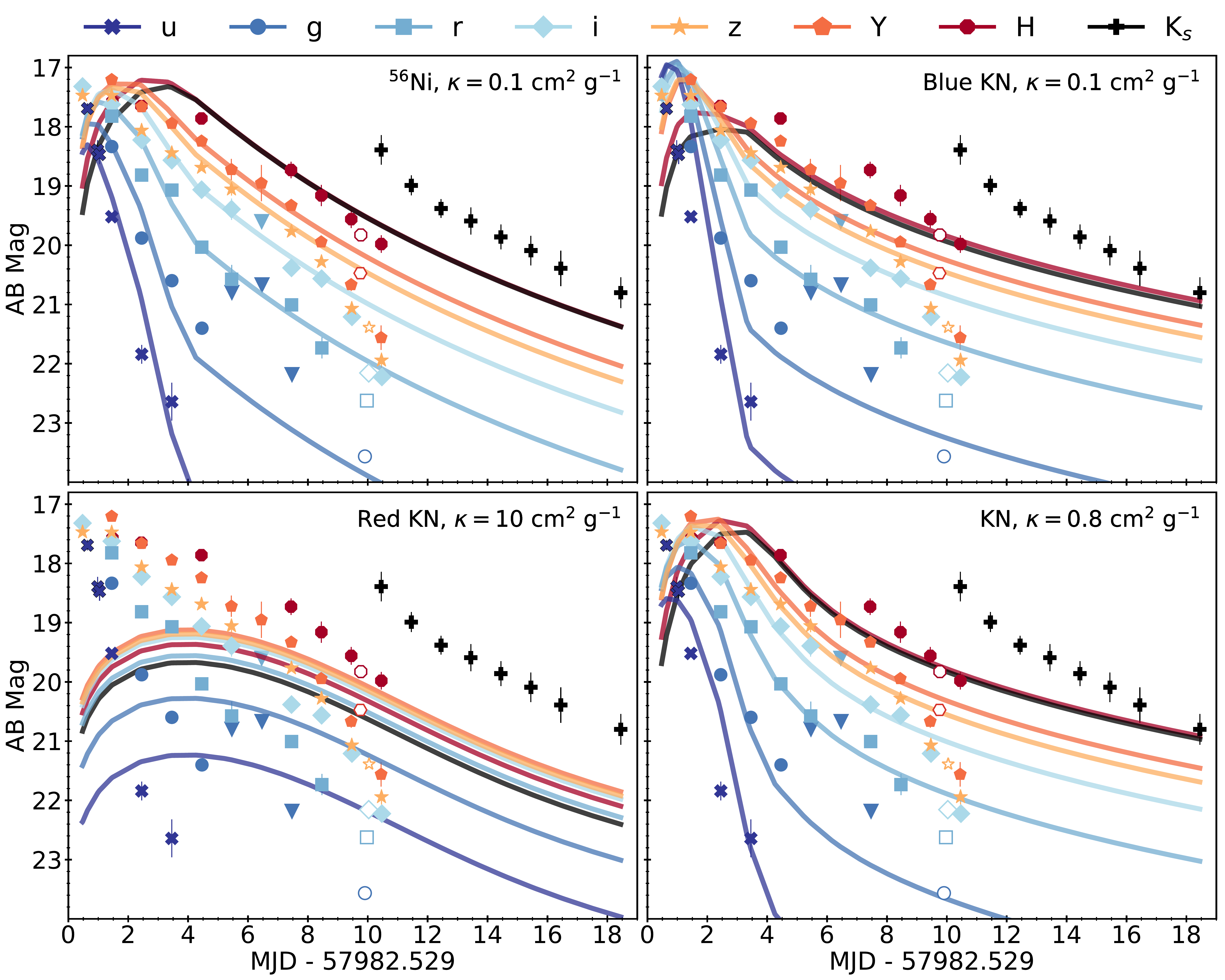}}
\caption{{\it Top Left:} Fitting the data with a Type I b/c SN model powered by the radioactive decay of $^{56}$Ni. This model clearly fails to capture the late time NIR behavior and requires an unphysically large fraction of the ejecta to be synthesized into nickel (\apx75\%). {\it Top Right:} Fitting the data with a single component ``blue'' KN model. Like the SN model, this fit is unable to capture the late time NIR behavior and overall spectra shape. {\it Bottom Left:} Fitting the data with a single component ``red'' KN model. This model clearly fails to capture any of the observed behavior. {\it Bottom Right:} Fitting the data with a single-component KN model with the opacity as a free parameter. Again, this model fails to capture the late time NIR behavior. This is suggestive of the fact that we need to model multiple ejecta components simultaneously. Error bars are given at the $1\sigma$ level in all panels, but may be smaller than the points.}
\label{fig:lc_failed}
\end{center}
\end{figure*}

We next turn to $r$-process heating, using the model outlined in \citet{Metzger2017} and implemented in \citet{villar2017}. This model includes the ejecta mass, ejecta velocity and opacity as fitted parameters (as well as the temperature floor and scatter). Within this context we first assume an Fe-peak opacity of $\kappa = 0.1$ cm$^{2}$ g$^{-1}$ (our "blue" model; e.g., as assumed historically in \citealt{li1998}) and fit for the ejecta mass and velocity.  This model, with $M_{\rm ej}\approx 0.03$ M$_\odot$ and $v_{\rm ej}\approx 0.18$c, adequately describes the early light curves ($\lesssim 3$ d), but again is a poor fit to the NIR light curves.  More recent calculations indicate that lanthanide-rich ejecta are expected to have a much higher opacity of $\kappa = 10$ cm$^{2}$ g$^{-1}$, leading to a ``red'' kilonova (e.g., \citealt{Barnes2013}).  However, such a model (our ``red" model), with best fit values $M_{\rm ej}\approx 0.03$ M$_\odot$ and $v_{\rm ej}\approx 0.27$c, produces a poor fit to the data as well.  In particular, the model light curves exhibit an initial rise for $\approx 4$ days, in contrast to the observed rapid decline at early time, especially in the UV and blue optical bands.  Finally, we allow the opacity to vary as a free parameter, finding a best fit value of $\kappa \approx 0.82$ cm$^{2}$ g$^{-1}$, and an associated $M_{\rm ej}\approx 0.04$ M$_\odot$ and $v_{\rm ej}\approx 0.27$c.  However, this model again fails to reproduce the initial rapid decline in the UV, as well as the NIR light curves.  We therefore conclude that $r$-process heating with a single value for the opacity cannot explain the observed light curve evolution and colors. The final light curves for these models can be seen in \autoref{fig:lc_failed}.

Inspired by the multi-component observed SED (Figure~\ref{fig:sed}) and by the failure of single-component models to capture both the early rapid decline and the late-time red colors, we explore two multi-component models: (i) a two-component ``blue'' ($\kappa = 0.5$ cm$^{2}$ g$^{-1}$) plus ``red'' ($\kappa$ as a free parameter) model; and (ii) a three-component ``blue'' ($\kappa = 0.5$ cm$^{2}$ g$^{-1}$) plus ``purple'' ($\kappa = 3$ cm$^{2}$ g$^{-1}$) plus ``red'' ($\kappa = 10$ cm$^{2}$ g$^{-1}$) model. These values were recently shown by \citet{Tanaka2017} to roughly capture the detailed opacity from radiative transfer simulations. For each component we leave $M_{\rm ej}$ and $v_{\rm ej}$ as free parameters.

First, we explore the two-component model (with eight free parameters); we vary the ejecta masses, ejecta velocities, and temperature floors, the red component opacity, and a single scatter term. We find that the ``blue'' component has $M_\mathrm{ej}^\mathrm{blue}\approx 0.01$ M$_\odot$ and $v_\mathrm{ej}^\mathrm{blue}\approx 0.27$c (with errors of roughly 10\%), in good agreement with our inference from the SED at early time (\autoref{sec:SED}). The ``red'' component has a much larger mass of $M_\mathrm{ej}^\mathrm{red}\approx 0.04$ M$_\odot$ but a slower velocity of $v_\mathrm{ej}^\mathrm{red}\approx 0.12$c. The best-fit opacity of this component is $\kappa\approx 3.3$ cm$^{2}$ g$^{-1}$, lower than expected for lanthanide-rich ejecta. We find that most of the parameters are uncorrelated, with the exception of the red component's opacity and ejecta velocity, which have a Pearson correlation coefficient of $\sim0.67$. The resulting parameters and uncertainties from the MCMC fitting are summarized in Table~\ref{tab:KN}.

For the three-component model (with ten free parameters) we find similar values for the ``blue'' component ($M_\mathrm{ej}^\mathrm{blue}\approx 0.01$ M$_\odot$ and $v_\mathrm{ej}^\mathrm{blue}\approx 0.27$c) and the ``purple'' component ($M_\mathrm{ej}^\mathrm{purple}\approx 0.03$ M$_\odot$ and $v_\mathrm{ej}^\mathrm{purple}\approx 0.11$c).  The ``red'' component is sub-dominant with $M_\mathrm{ej}^\mathrm{red}\approx 0.01$ M$_\odot$ and $v_\mathrm{ej}^\mathrm{red}\approx 0.16$c); see Table~\ref{tab:KN}. These ejecta parameters are consistent with those determined from independent modeling of the optical and NIR spectra \citep{DECamPaper4,DECamPaper3}.

Both sets of models are shown in Figure~\ref{fig:lc_good} and are essentially indistinguishable.  Both provide a much better fit to the data than the single-component models described above, capturing both the initial blue colors and rapid decline, as well as the later redder colors and NIR light curves.  Their similar WAIC scores suggest that neither model is statistically preferred. The two models differ most drastically at $\lesssim 5$ days in the $K_s$-band, where the two-component model is double-peaked, while the three-component model is single peaked. While neither model fully captures every feature of the light curves, it is remarkable that these simplified semi-analytic models produce such high quality fits over a wide range of wavelength and time. 

\section{Implications}
\label{sec:interpretation}

In the multi-component models, we can interpret each component as arising from distinct physical regions within the merger ejecta. In both models, the high velocity of the blue KN ejecta suggests that it originates from the shock-heated polar region created when the neutron stars collide (e.g.~\citealt{Oechslin2007,Bauswein2013,Sekiguchi2016}). This dominant blue component is also seen in early time optical spectra \citep{DECamPaper3}. By contrast, the low velocity red KN component in our three-component model could originate from the dynamically-ejected tidal tails in the equatorial plane of the binary (e.g., \citealt{Rosswog1999,Hotokezaka2013}), in which case the relatively high ejecta mass $\approx 0.01$ M$_{\odot}$ suggests an asymmetric mass ratio of the merging binary (q$\lesssim0.8$; \citealt{Hotokezaka2013}).

In both multi-component models we find that the $\kappa\approx 3$ cm$^{2}$ g$^{-1}$ ejecta dominates by mass.  The lower velocity of this component suggests an origin in the post-merger accretion disk outflow.  Our inferred ejecta mass is consistent with that expected for a massive $\sim 0.1$ M$_{\odot}$ torus (e.g., \citealt{Just2015,Siegel&Metzger2017}).  Similarly, the disk outflow composition is predicted to be dominated by $Y_{e}\sim 0.3$ matter that produces the $\kappa\approx 3$ cm$^{2}$ g$^{-1}$ component of the KN emission \citep{Tanaka2017} as we observe. The fitted opacity indicates that the hyper-massive neutron star remnant is relatively short-lived ($\sim 30$ ms; \citealt{Fernandez2013,Just2015,kasen15}). We additionally find that in both models the total kinetic energy is roughly $(1-2)\times 10^{51}$ erg.

The fact that our multi-component models fit the data well provides strong evidence for the production of both light and heavy $r$-process nuclei, addressing one of the long-standing mysteries in astrophysics \citep{Burbidge1957,Cameron1957}. We quantify this statement by comparing our blue and red ejecta masses to those necessary to reproduce the Milky Way $r$-process production rate.  For heavy $r$-process elements (red KN), the Milky Way inferred production rate is $\dot{M}_\mathrm{rp,A\gtrsim 140}\approx 10^{-7}$ M$_\odot$ yr$^{-1}$ \citep{bauswein2014nucleosynthesis}. For light $r$-process elements (blue KN), the production rate is $\dot{M}_\mathrm{rp,A\gtrsim 100}\approx 7\times 10^{-7}$ M$_\odot$ yr$^{-1}$ \citep{qian2000supernovae}. Using a conservative estimate on the local BNS merger rate estimated by \citet{abbott2016upper}, $R_\mathrm{0}\approx 1000$ Gpc$^{-3}$ yr$^{-1}$, and a volume density of Milky Way-like galaxies of $\approx 0.01$ Mpc$^{-3}$, we estimate the Milky Way rate of KN as $R_\mathrm{MW}\approx 100$ Myr$^{-1}$.  Using this MW rate, we find that the average ejecta mass for a red KN is $\dot{M}_\mathrm{rp,A\gtrsim 140}/R_\mathrm{MW}\approx 0.001$ M$_\odot$ and for a blue KN it is $\dot{M}_\mathrm{rp,A\gtrsim 100}/R_\mathrm{MW}\approx 0.007$ M$_\odot$. These order-of-magnitude estimates are  smaller than our inferred ejecta masses for this event, although the discrepancy can potentially be mitigated when properly taking into account the fraction of $r$-process materials which remains in a gas phase in the ISM and galactic halo. Nevertheless, this exercise suggests that BNS mergers can reproduce the $r$-process yields found in the Milky Way and may be a dominant source of cosmic $r$-process nucleosynthesis.

\section{Discussions and Conclusions}
\label{sec:conc}

We presented a comprehensive UV, optical, and NIR data set for the first electromagnetic counterpart to be associated with a gravitational wave event.  Analysis of these data reveal that the emission is due to an $r$-process powered kilonova consisting of both ``blue'' and ``purple/red'' ejecta components. Models with $^{56}$Ni heating, Fe-peak opacities, or a single component of $r$-process opacity fail to match the observations.

Our models indicate that the total ejecta mass is $\approx 0.05$ M$_\odot$, with a high velocity ($v\approx 0.3c$) blue component and a slower ($v\approx 0.1-0.2c$) purple/red component.  The presence of both components and the relatively large ejecta mass suggests that binary neutron star mergers (like GW170817) dominate the cosmic $r$-process nucleosynthesis.

The data presented in this paper (and others in this series) represent by far the best observations of an $r$-process powered kilonova, and it is remarkable how well the observations match theoretical models.  This event also marks the true beginning of joint GW-EM multi-messenger astronomy.  We expect that this event will serve as a benchmark for future efforts to model and understand the behavior of these transients, and for the first time allow the development of data-driven kilonova models. The next Advanced LIGO/Virgo observing run (starting in Fall 2018) is expected to detect many more BNS events \citep{LIGOobs}. Follow-up of these events will provide further understanding of the ubiquity of the features seen in this event, the relationship between event and host properties and place even stronger constraints on $r$-process enrichment from BNS mergers.

{\acknowledgments
The Berger Time-Domain Group at Harvard is supported in part by the NSF through grants AST-1411763 and AST-1714498, and by NASA through grants NNX15AE50G and NNX16AC22G. PSC is grateful for support provided by the NSF through the Graduate Research Fellowship Program, grant DGE 1144152. VAV acknowledges support by the National Science Foundation through a Graduate Research Fellowship. DAB is supposed by NSF award PHY-1707954. We thank the University of Copenhagen, DARK Cosmology Centre, and the Niels Bohr International Academy for hosting R.J.F.\ during the discovery of GW170817/SSS17a, where he was participating in the Kavli Summer Program in Astrophysics, ``Astrophysics with gravitational wave detections."  This program was supported by the Kavli Foundation, Danish National Research Foundation, the Niels Bohr International Academy, and the DARK Cosmology Centre. The UCSC group is supported in part by NSF grant AST--1518052, the Gordon \& Betty Moore Foundation, the Heising-Simons Foundation, generous donations from many individuals through a UCSC Giving Day grant, and from fellowships from the Alfred P.\ Sloan Foundation and the David and Lucile Packard Foundation to R.J.F. EB acknowledges financial support from the European Research Council (ERC-StG-335936, CLUSTERS). DK \& EQ were funded in part by the Gordon and Betty Moore Foundation through Grant GBMF5076.

We are grateful for the heroic efforts of the entire staff at Gemini-South to continue obtaining observations of this target in evening twilight at high airmass as the object was setting.

This project used data obtained with the Dark Energy Camera (DECam), which was constructed by the Dark Energy Survey (DES) collaboration. Funding for the DES Projects has been provided by the DOE and NSF (USA), MISE (Spain), STFC (UK), HEFCE (UK), NCSA (UIUC), KICP (U. Chicago), CCAPP (Ohio State), MIFPA (Texas A\&M), CNPQ, FAPERJ, FINEP (Brazil), MINECO (Spain), DFG (Germany) and the collaborating institutions in the Dark Energy Survey, which are Argonne Lab, UC Santa Cruz, University of Cambridge, CIEMAT-Madrid, University of Chicago, University College London, DES-Brazil Consortium, University of Edinburgh, ETH Z{\"u}rich, Fermilab, University of Illinois, ICE (IEEC-CSIC), IFAE Barcelona, Lawrence Berkeley Lab, LMU M{\"u}nchen and the associated Excellence Cluster Universe, University of Michigan, NOAO, University of Nottingham, Ohio State University, University of Pennsylvania, University of Portsmouth, SLAC National Lab, Stanford University, University of Sussex, and Texas A\&M University.

Based in part on observations at Cerro Tololo Inter-American Observatory, National Optical Astronomy Observatory (NOAO Prop. ID: 2017B-0110, PI: E. Berger), which is operated by the Association of Universities for Research in Astronomy (AURA) under a cooperative agreement with the National Science Foundation

Based in part on observations obtained at the Gemini Observatory (Program IDs GS-2017B-Q-8 and GS-2017B-DD-4; PI: Chornock), which is operated by the Association of Universities for Research in Astronomy, Inc., under a cooperative agreement with the NSF on behalf of the Gemini partnership: the National Science Foundation (United States), the National Research Council (Canada), CONICYT (Chile), Ministerio de Ciencia, Tecnolog\'{i}a e Innovaci\'{o}n Productiva (Argentina), and Minist\'{e}rio da Ci\^{e}ncia, Tecnologia e Inova\c{c}\~{a}o (Brazil).}

\section{Summary of Photometry}
\label{app:photometry}

\startlongtable
\begin{deluxetable}{lcccc}
\tabletypesize{\footnotesize}
\tablecolumns{5}
\tablewidth{0pt}
\tablecaption{Summary of Photometry 
	          \label{tab:phot}}
\tablehead{
\colhead{Telescope} &
\colhead{Instrument} &
\colhead{Filter} & 
\colhead{MJD} &
\colhead{Mag [AB]}
}
\startdata
Blanco & DECam & $i$ & 0.4745 & $17.48 \pm 0.03$ \\
Blanco & DECam & $z$ & 0.4752 & $17.59 \pm 0.03$ \\
{\it Swift} & UVOT & $M2$ & 0.627 & $21.14 \pm 0.23$ \\
{\it Swift} & UVOT & $W1$ & 0.634 & $19.53 \pm 0.12$ \\
{\it Swift} & UVOT & $U$ & 0.639 & $18.20 \pm 0.09$ \\
{\it Swift} & UVOT & $W2$ & 0.643 & $20.76 \pm 0.20$ \\
{\it Swift} & UVOT & $U$ & 0.981 & $18.90 \pm 0.17$ \\
{\it Swift} & UVOT & $U$ & 1.043 & $18.98 \pm 0.16$ \\
Blanco & DECam & $Y$ & 1.4478 & $17.32 \pm 0.03$ \\
Blanco & DECam & $z$ & 1.4485 & $17.59 \pm 0.02$ \\
Blanco & DECam & $i$ & 1.4492 & $17.78 \pm 0.02$ \\
Blanco & DECam & $r$ & 1.4499 & $18.04 \pm 0.02$ \\
Blanco & DECam & $g$ & 1.4506 & $18.66 \pm 0.03$ \\
Blanco & DECam & $u$ & 1.4512 & $19.94 \pm 0.05$ \\
Gemini-South & FLAMINGOS-2 & $H$ & 1.471 & $17.63 \pm 0.10$ \\
Gemini-South & FLAMINGOS-2 & $H$ & 2.439 & $17.71 \pm 0.09$ \\
Blanco & DECam & $Y$ & 2.4461 & $17.77 \pm 0.03$ \\
Blanco & DECam & $z$ & 2.4472 & $18.18 \pm 0.03$ \\
Blanco & DECam & $i$ & 2.4479 & $18.38 \pm 0.03$ \\
Blanco & DECam & $r$ & 2.4486 & $19.03 \pm 0.03$ \\
Blanco & DECam & $g$ & 2.4492 & $20.21 \pm 0.05$ \\
Blanco & DECam & $u$ & 2.4515 & $22.26 \pm 0.16$ \\
Blanco & DECam & $Y$ & 3.4541 & $18.05 \pm 0.03$ \\
Blanco & DECam & $z$ & 3.4551 & $18.56 \pm 0.03$ \\
Blanco & DECam & $u$ & 3.4556 & $23.06 \pm 0.32$ \\
Blanco & DECam & $i$ & 3.4558 & $18.73 \pm 0.03$ \\
Blanco & DECam & $r$ & 3.4564 & $19.29 \pm 0.04$ \\
Blanco & DECam & $g$ & 3.4571 & $20.93 \pm 0.08$ \\
Gemini-South & FLAMINGOS-2 & $H$ & 4.445 & $17.92 \pm 0.10$ \\
Blanco & DECam & $Y$ & 4.4467 & $18.35 \pm 0.03$ \\
Blanco & DECam & $z$ & 4.4491 & $18.81 \pm 0.03$ \\
Blanco & DECam & $i$ & 4.4516 & $19.22 \pm 0.03$ \\
Blanco & DECam & $r$ & 4.4552 & $20.25 \pm 0.05$ \\
Blanco & DECam & $g$ & 4.4624 & $21.73 \pm 0.11$ \\
Blanco & DECam & $Y$ & 5.4460 & $18.83 \pm 0.18$ \\
Blanco & DECam & $z$ & 5.4484 & $19.17 \pm 0.11$ \\
Blanco & DECam & $i$ & 5.4508 & $19.55 \pm 0.18$ \\
Blanco & DECam & $r$ & 5.4545 & $20.79 \pm 0.24$ \\
Blanco & DECam & $g$ & 5.462 & $>20.80$ \\
Blanco & DECam & $Y$ & 6.4458 & $19.06 \pm 0.31$ \\
Blanco & DECam & $r$ & 6.457 & $>19.60$ \\
Blanco & DECam & $g$ & 6.468 & $>20.67$ \\
Gemini-South & FLAMINGOS-2 & $H$ & 7.438 & $18.79 \pm 0.14$ \\
Blanco & DECam & $Y$ & 7.4448 & $19.44 \pm 0.05$ \\
Blanco & DECam & $z$ & 7.4509 & $19.89 \pm 0.05$ \\
Blanco & DECam & $i$ & 7.4533 & $20.54 \pm 0.05$ \\
Blanco & DECam & $r$ & 7.4581 & $21.23 \pm 0.11$ \\
Blanco & DECam & $g$ & 7.469 & $>22.19$ \\
Blanco & DECam & $Y$ & 8.4446 & $20.06 \pm 0.07$ \\
Gemini-South & FLAMINGOS-2 & $H$ & 8.452 & $19.22 \pm 0.18$ \\
Blanco & DECam & $z$ & 8.4543 & $20.40 \pm 0.06$ \\
Blanco & DECam & $i$ & 8.4591 & $20.72 \pm 0.06$ \\
Blanco & DECam & $r$ & 8.4688 & $21.95 \pm 0.18$ \\
Blanco & DECam & $Y$ & 9.4457 & $20.78 \pm 0.11$ \\
Gemini-South & FLAMINGOS-2 & $H$ & 9.449 & $19.62 \pm 0.15$ \\
Blanco & DECam & $z$ & 9.4659 & $21.19 \pm 0.07$ \\
Blanco & DECam & $i$ & 9.4712 & $21.37 \pm 0.06$ \\
{\it HST} & WFC3/IR & F110W & 9.753 & $20.57 \pm 0.04$ \\
{\it HST} & WFC3/IR & F160W & 9.768 & $19.89 \pm 0.04$ \\
{\it HST} & WFC3/UVIS & F336W & 9.819 & $26.92 \pm 0.27$ \\
{\it HST} & ACS/WFC & F475W & 9.905 & $23.95 \pm 0.06$ \\
{\it HST} & ACS/WFC  & F625W & 9.969 & $22.88 \pm 0.07$ \\
{\it HST} & ACS/WFC  & F775W & 10.032 & $22.35 \pm 0.08$ \\
{\it HST} & ACS/WFC  & F850LP & 10.045 & $21.53 \pm 0.05$ \\
Blanco & DECam & $Y$ & 10.4462 & $21.67 \pm 0.21$ \\
Gemini-South & FLAMINGOS-2 & $K_s$ & 10.449 & $18.43 \pm 0.25$ \\
Gemini-South & FLAMINGOS-2 & $H$ & 10.453 & $20.04 \pm 0.15$ \\
Blanco & DECam & $z$ & 10.4583 & $22.06 \pm 0.13$ \\
Blanco & DECam & $i$ & 10.4715 & $22.38 \pm 0.10$ \\
Gemini-South & FLAMINGOS-2 & $K_s$ & 11.455 & $19.03 \pm 0.17$ \\
Gemini-South & FLAMINGOS-2 & $K_s$ & 12.447 & $19.42 \pm 0.16$ \\
Gemini-South & FLAMINGOS-2 & $K_s$ & 13.441 & $19.63 \pm 0.23$ \\
Gemini-South & FLAMINGOS-2 & $K_s$ & 14.446 & $19.90 \pm 0.21$ \\
Gemini-South & FLAMINGOS-2 & $K_s$ & 15.447 & $20.13 \pm 0.25$ \\
Gemini-South & FLAMINGOS-2 & $K_s$ & 16.446 & $20.43 \pm 0.30$ \\
Gemini-South & FLAMINGOS-2 & $K_s$ & 18.450 & $20.84 \pm 0.26$ \\
\enddata
\tablecomments{Summary of photometry from \autoref{sec:observations}. Dates are give in days relative to the time of the GW trigger (MJD = 57982.529). Photometry {\it is not} corrected for extinction. Limits are given at the $3\sigma$ level. Error bars are given at the $1\sigma$ level.}
\end{deluxetable}

\bibliographystyle{yahapj.bst}
\bibliography{references.bib}

\end{document}